\documentclass{elsart}
\usepackage{amssymb}
\usepackage{epsfig}
\usepackage{graphics}
\usepackage{color}
\definecolor{mygreen}{rgb}{0,.8,0}

\def\ifm#1{\relax\ifmmode#1\else$#1$\fi}  
  \def\ab{\ifm{\sim}}  \def\x{\ifm{\times}}
\def\f{\ifm{\phi}}  \def\DAF{DA\char8NE}   
\def\pic{\ifm{\pi^+\pi^-}}  \def\to{\ifm{\rightarrow}}
  \def\dif{\hbox{d\kern.5mm}}
\def\up#1{\ifm{^{#1}}}  \def\deg{\ifm{^\circ}}
   \def\gam{\ifm{\gamma}}
  
\def\epm{\ifm{e^+e^-}}  
\def\dn#1{\ifm{_{#1}}}     
\def\eps{\ifm{\epsilon}}   
\let\cal=\mathcal   \def\ORD#1!{\ifm{{\cal O}\hbox{(#1)}}}
\def\pt#1,#2,{\ifm{#1\x10^{#2}}} 
\def\plm{\ifm{\pm}}

\def\ifm#1{\relax\ifmmode#1\else$#1$\fi}
\def\DAF{DA\char8NE}  
\def\f{\ifm{\phi}}   \def\epm{\ifm{e^+e^-}}
\def\ab{\ifm{\sim}}  \def\x{\ifm{\times}}
\def\gam{\ifm{\gamma}}  \def\pic{\ifm{\pi^+\pi^-}}
\def\pt#1,#2,{\ifm{#1\x10^{#2}}}  \def\dif{\ifm{{\rm d}\,}}
\renewcommand{\to}{\ensuremath{\rightarrow}}

\def\up#1{$^{#1}$}  \def\dn#1{$_{#1}$}  
   \def\plm{\ifm{\pm}}

\makeatletter
\newdimen\z@ \z@=0pt 
\newskip\z@skip \z@skip=0pt plus0pt minus0pt
\def\m@th{\mathsurround=\z@}
\def\ialign{\everycr{}\tabskip\z@skip\halign} 
\def\eqalign#1{\null\,\vcenter{\openup\jot\m@th
  \ialign{\strut\hfil$\displaystyle{##}$&$\displaystyle{{}##}$\hfil
      \crcr#1\crcr}}\,}
\makeatother

\newcommand{\aff}[2]{Dipartimento di Fisica dell'Universit\`a #1 e Sezione INFN, #2, Italy.}
\newcommand{\affd}[1]{Dipartimento di Fisica dell'Universit\`a e Sezione INFN, #1, Italy.}
\begin{document}
\begin{frontmatter}

\title{
Upper limit on the $\eta \rightarrow \pi^+\pi^-$ branching ratio with the KLOE
detector.
}
\collab{The KLOE Collaboration}

\author[Na]{F.~Ambrosino},
\author[Frascati]{A.~Antonelli},
\author[Frascati]{M.~Antonelli},
\author[Roma3]{C.~Bacci},
\author[Roma3]{M.~Barva},
\author[Frascati]{P.~Beltrame},
\author[Frascati]{G.~Bencivenni},
\author[Frascati]{S.~Bertolucci},
\author[Roma1]{C.~Bini},
\author[Frascati]{C.~Bloise},
\author[Roma1]{V.~Bocci},
\author[Frascati]{F.~Bossi},
\author[Frascati,Virginia]{D.~Bowring},
\author[Roma3]{P.~Branchini},
\author[Roma1]{R.~Caloi},
\author[Frascati]{P.~Campana},
\author[Frascati]{G.~Capon},
\author[Na]{T.~Capussela},
\author[Roma2]{G.~Carboni},
\author[Roma3]{F.~Ceradini},
\author[Pisa]{F.~Cervelli},
\author[Frascati]{S.~Chi},
\author[Na]{G.~Chiefari},
\author[Frascati]{P.~Ciambrone},
\author[Virginia]{S.~Conetti},
\author[Roma1]{E.~De~Lucia},
\author[Roma1]{A.~De~Santis},
\author[Frascati]{P.~De~Simone},
\author[Roma1]{G.~De~Zorzi},
\author[Frascati]{S.~Dell'Agnello},
\author[Karlsruhe]{A.~Denig},
\author[Roma1]{A.~Di~Domenico},
\author[Na]{C.~Di~Donato},
\author[Pisa]{S.~Di~Falco},
\author[Roma3]{B.~Di~Micco},
\author[Na]{A.~Doria},
\author[Frascati]{M.~Dreucci},
\author[Roma3]{A.~Farilla},
\author[Frascati]{G.~Felici},
\author[Frascati]{A.~Ferrari},
\author[Frascati]{M.~L.~Ferrer},
\author[Frascati]{G.~Finocchiaro},
\author[Frascati]{C.~Forti},
\author[Roma1]{P.~Franzini},
\author[Roma1]{C.~Gatti},
\author[Roma1]{P.~Gauzzi},
\author[Frascati]{S.~Giovannella},
\author[Lecce]{E.~Gorini},
\author[Roma3]{E.~Graziani},
\author[Pisa]{M.~Incagli},
\author[Karlsruhe]{W.~Kluge},
\author[Moscow]{V.~Kulikov},
\author[Roma1]{F.~Lacava},
\author[Frascati]{G.~Lanfranchi},
\author[Frascati,StonyBrook]{J.~Lee-Franzini},
\author[Roma1]{D.~Leone},
\author[Frascati]{M.~Martemianov},
\author[Frascati]{M.~Martini},
\author[Na]{P.~Massarotti},
\author[Frascati]{M.~Matsyuk},
\author[Frascati]{W.~Mei},
\author[Na]{S.~Meola},
\author[Roma2]{R.~Messi},
\author[Frascati]{S.~Miscetti},
\author[Frascati]{M.~Moulson},
\author[Karlsruhe]{S.~M\"uller},
\author[Frascati]{F.~Murtas},
\author[Na]{M.~Napolitano},
\author[Roma3]{F.~Nguyen},
\author[Frascati]{M.~Palutan},
\author[Roma1]{E.~Pasqualucci},
\author[Frascati]{L.~Passalacqua},
\author[Roma3]{A.~Passeri},
\author[Frascati,Energ]{V.~Patera},
\author[Na]{F.~Perfetto},
\author[Roma1]{L.~Pontecorvo},
\author[Lecce]{M.~Primavera},
\author[Frascati]{P.~Santangelo},
\author[Roma2]{E.~Santovetti},
\author[Na]{G.~Saracino},
\author[Frascati]{B.~Sciascia},
\author[Frascati,Energ]{A.~Sciubba},
\author[Pisa]{F.~Scuri},
\author[Frascati]{I.~Sfiligoi},
\author[Frascati]{T.~Spadaro},
\author[Roma3]{E.~Spiriti},
\author[Roma1]{M.~Testa},
\author[Roma3]{L.~Tortora},
\author[Frascati]{P.~Valente},
\author[Karlsruhe]{B.~Valeriani},
\author[Frascati]{G.~Venanzoni},
\author[Roma1]{S.~Veneziano},
\author[Lecce]{A.~Ventura},
\author[Roma1]{S.~Ventura},
\author[Roma3]{R.~Versaci},
\author[Na]{I.~Villella},
\author[Beijing,Frascati]{G.~Xu},

\address[Beijing]{Institute of High Energy 
Physics of Academica Sinica, Beijing, China.}
\address[Frascati]{Laboratori Nazionali di Frascati dell'INFN, 
Frascati, Italy.}
\address[Karlsruhe]{Institut f\"ur Experimentelle Kernphysik, 
Universit\"at Karlsruhe, Germany.}
\address[Lecce]{\affd{Lecce}}
\address[Moscow]{Institute for Theoretical 
and Experimental Physics, Moscow, Russia.}
\address[Na]{Dipartimento di Scienze Fisiche dell'Universit\`a 
``Federico II'' e Sezione INFN,
Napoli, Italy}
\address[Pisa]{\affd{Pisa}}
\address[Energ]{Dipartimento di Energetica dell'Universit\`a 
``La Sapienza'', Roma, Italy.}
\address[Roma1]{\aff{``La Sapienza''}{Roma}}
\address[Roma2]{\aff{``Tor Vergata''}{Roma}}
\address[Roma3]{\aff{``Roma Tre''}{Roma}}
\address[StonyBrook]{Physics Department, State University of New 
York at Stony Brook, USA.}
\address[Virginia]{Physics Department, University of Virginia, USA.}
{\small Corresponding author: Cesare Bini, e-mail
  cesare.bini@roma1.infn.it, tel +390649914266, fax +39064957697}
\begin{abstract}
We have searched with the KLOE detector for the $P$ and $CP$ violating decay 
$\eta\rightarrow\pi^+\pi^-$ in a sample of \pt1.55,7, $\eta$'s from the decay 
$\phi\rightarrow\eta\gamma$ of \f-mesons produced in \epm\ annihilations at \DAF.
No signal is found. We obtain the upper limit BR($\eta\rightarrow\pi^+\pi^-$)$<$
1.3$\times 10^{-5}$ at 90\% confidence level.
\end{abstract}
\begin{keyword}Decays of $\eta$ mesons, discrete symmetries
\PACS 11.30.Er, 13.25.Jk, 14.40.Aq
\end{keyword}
\end{frontmatter}

The study of $\eta$ decays provides an excellent laboratory for testing the 
validity of symmetries of the physical world. The decay $\eta 
\rightarrow \pi^+\pi^-$ violates both $P$ and $CP$ invariance. 
In the Standard Model, the reaction can proceed only via the weak interaction 
with a branching ratio of order 10\up{-27} according to
Ref.\cite{Shab}. Higher branching ratios are conceivable
either introducing a CP violation term in the QCD lagrangian through the
so-called $\theta$ term (a branching ratio up to 10\up{-17} can be obtained in
this scheme compatible with the experimental limit on the neutron electric dipole momentum)
or allowing a CP violation in the extended Higgs sector (in this case 
10\up{-15} can be reached) as described in Ref.\cite{Shab}.  
The detection of this decay at any level accessible today would signal 
$P$ and $CP$ violation from new sources, beyond any considerable extension of the 
Standard Model. The best published previous result from a direct search for the
$\eta$ decay
to $\pi^+\pi^-$ was obtained in 1999: 
$BR(\eta \rightarrow \pi^+\pi^-)< 3.3\times 10^{-4}$ at 90\%
confidence level \cite{russilimite}.

We present here the result of a direct search for this decay performed with the KLOE experiment at the
Frascati $\phi$ factory, \DAF, based on an integrated luminosity of 350 pb$^{-1}$ collected in the years 2001
and 2002.
\DAF\ is an $e^+e^-$ collider working at the centre of mass energy of 1019.5 MeV,
i.e., at the $\phi$ mass. $\phi$ mesons are produced nearly at rest in the
laboratory. Beams in \DAF\ collide with a crossing angle of $\pi-0.025$ 
rad. $\phi$-mesons are therefore produced with a momentum 
$|\vec{p_{\phi}}|\sim 12.5$ MeV/c in the horizontal plane, directed towards 
the centre of the storage rings. The precise value of the $\phi$ momentum, 
together with the centre of mass energy and the beam spot position, is 
determined run by run using Bhabha scattering events.

The total cross section for \epm\to\f\ is \ab3 $\mu$b. $\eta$ mesons are
copiously produced at a typical 
rate of $\sim$2 Hz through the radiative decay $\phi\rightarrow\eta\gamma$,
which has a branching ratio of 
1.3\%.
In the decay chain $\phi\rightarrow\eta\gamma\rightarrow\pi^+\pi^-\gamma$, the
photon is monochromatic 
($E_\gamma=363$ MeV) and is emitted according to an 
angular distribution $dN/d\cos\theta_{\gamma}\sim(1+\cos^2\theta_{\gamma})$, 
where $\theta_\gamma$ is the polar angle of the emitted photon with respect to
the beam line. The invariant 
mass of the $\pi^+\pi^-$ pair is equal to the $\eta$ mass: $M_{\pi\pi}$=$M_{\eta}$=547.3 MeV.

The KLOE detector consists of
a large-volume drift chamber \cite{dch} (3.3 m length
and 2 m radius), operated with a 90\% helium-10\% isobutane gas mixture, and
a sampling calorimeter \cite{calo} made of lead and
scintillating fibres. The calorimeter consists of a cylindrical barrel and 
two endcaps providing a solid angle coverage of 98\%. A superconducting coil 
surrounds the entire detector and produces a solenoidal field $B$=0.52 T.

Tracks are reconstructed in the drift chamber with a momentum resolution
of $\sigma(p_\perp)/p_\perp < $ 0.4\%. Clusters of energy deposits in
the calorimeter are classified either as associated to charged tracks 
(charged pions, electrons or muons) or as isolated (photons, $K_L$). 
Photon energies and arrival times are measured with
resolutions of $\sigma_E/E=5.7\%/\sqrt{E({\rm GeV})}$ and 
$\sigma_t=54{\rm ps}/\sqrt{E({\rm GeV})}\oplus 50$ ps; impact positions are
measured with a resolution
of a few centimetres. The readout
granularity is 4.5 $\times$ 4.5 cm$^2$ in the plane transverse to the fibres,
and is segmented in five layers along the particles direction.
The trigger \cite{TRGnim} is based on the detection of at least
two energy deposits in the
calorimeter above a threshold that ranges between 50 MeV in the barrel 
and 150 MeV in the endcaps. The higher machine background rates at small angle
requires a higher threshold in the endcaps.

Samples of simulated events are obtained using the GEANFI \cite{NIMoffline} code based on GEANT: 
event generators for any specific final state, including decay dynamics and radiative corrections, 
are provided together with the detailed description of the geometry and
the response of each sub-detector.

The decay chain $\phi\rightarrow\eta\gamma\rightarrow\pi^+\pi^-\gamma$
is searched for by selecting events with two tracks of opposite charge with a
vertex near the $e^+e^-$
interaction point and one prompt
photon matching the missing energy and
momentum obtained from the $\pi^+\pi^-$ pair and the $\phi$ kinematic 
variables. The vertex is required to be inside a
cylinder 20 cm long and 8 cm in radius, with axis parallel to the beam line, centred at the
beam spot position. The polar angle $\theta_t$ of each
track is required to satisfy 45\deg$<\theta_t<$135\deg. 
A prompt photon is detected as an energy cluster not associated to any track, with 
time of flight $T_{cl}$, distance from the beam spot position $R_{cl}$, and energy $E_{cl}$  
satisfying the condition $|T_{cl}-R_{cl}/c|<5\sigma_t(E_{cl})$, where
$\sigma_t$ is the energy-dependent time resolution. In order to match
the missing energy and momentum obtained from the $\pi^+\pi^-$ pair with 
the photon kinematics, the angle $\psi$ between the direction of the 
missing momentum and the direction of the photon is required to be less 
than 0.15 rad.

\begin{figure}[ht]
\centering
\vspace{2.cm}
\epsfig{file=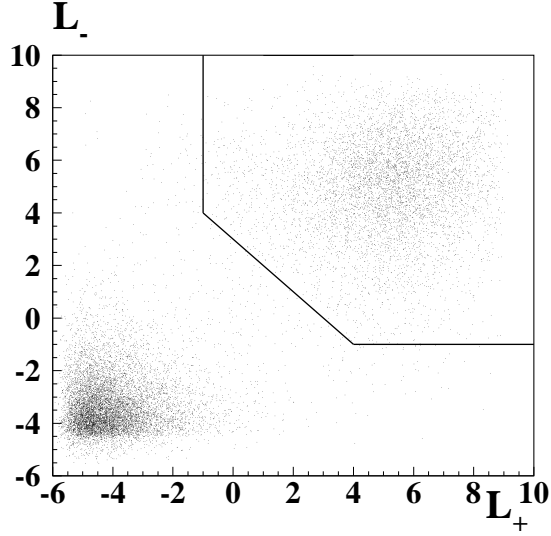,width=8cm}
\caption{\small{ Scatter plot of the
likelihood variable for positive (L\dn+) and negative (L\dn-) particles in
arbitrary units. 
The line is the cut applied to select pion and muon pairs (above) from electron
pairs (below).
}}
\label{cut1}
\end{figure}

Each track is extrapolated to the calorimeter and is required to
be associated with a calorimeter cluster. 
A major source of background is due to radiative Bhabha events, 
\epm\to\epm\gam. 
Rejection of these events is based on the shower energy deposition in the
calorimeter, 
on the time of flight, which 
is different for electrons and pions, and on kinematics. 
A likelihood estimator 
is constructed using the following information: the total energy of the
cluster and the maximum energy release among the five planes of the
calorimeter; the energy release in the first and in the last fired calorimeter
plane; and $|T_{cl}-L/c|$, where $L$ is the track length
from the interaction point to the centroid of the cluster. The probability
density function for each variable is obtained using samples
of unambiguously identified 
pions from $\pi^+\pi^-\pi^0$ and $\pi^+\pi^-$ events.
The separation between $\pi^+\pi^-\gamma$ and $e^+e^-\gamma$ events based on
the values of the likelihood estimators L\dn+ and L\dn- for 
positive and negative
particles is shown
in Fig. \ref{cut1}. 

\begin{figure}[ht]
  \centering
   \epsfig{file=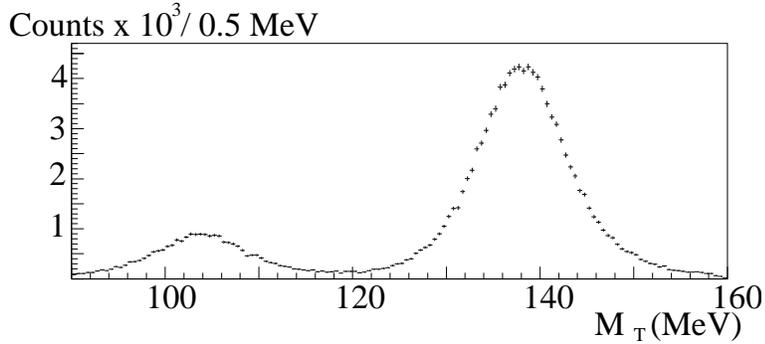,width=10cm}
    \caption{Distribution of the {\it track mass} variable $M_T$ showing the separation between the peaks due to $\pi^+\pi^-\gamma$ events (right peak) and to $\mu^+\mu^-\gamma$ events (left peak).}
  \label{cut2}
\end{figure}

$\mu^+\mu^-\gamma$ and residual $e^+e^-\gamma$ events are rejected using 
the so called {\it track-mass} variable $M_T$. 
$M_T$ is the particle mass computed assuming the \f\ decays to two particles
of identical mass plus a photon. $M_T$ is given by:
\begin{equation}
|\vec{p}_{\phi}-\vec{p}_1-\vec{p}_2|=
E_{\phi}-\sqrt{p_1^2+M_T^2}-\sqrt{p_2^2+M_T^2}
\end{equation}
where $\vec{p}_1$ and $\vec{p}_2$ are the three-momenta of the two pions and
$E_{\phi}$ and $\vec{p}_{\phi}$ are the total centre-of-mass energy and 
momentum respectively.
The track mass value is obtained from tracking information only and is very 
weakly correlated to the likelihood estimator value. The requirement 
129 $<M_T<$ 149 MeV selects \pic\gam\ events; see Fig. \ref{cut2}.
This cut, together with the cut on $\psi$ described above, ensures that 
background due to $\phi\rightarrow \pi^+\pi^-\pi^0$ 
events is negligible.

The $M_{\pi\pi}$ spectrum of the selected events ranges from $2m_{\pi}$=279
MeV to $M_{\phi}$=1019.5 MeV. Apart from the hypothetical signal, the physical
processes which give $\pi^+\pi^-\gamma$ final states are
$e^+e^-\rightarrow\pi^+\pi^-$ accompanied by ISR (initial state radiation) or
FSR (final state radiation), $\phi\rightarrow f_0(980)\gamma$ with
$f_0(980)\rightarrow \pi^+\pi^-$ and $\phi\rightarrow\rho^{\pm}\pi^{\pm}$ with
$\rho^{\pm}\rightarrow\pi^{\pm}\gamma$. Analyses of the full spectrum in
different photon angular ranges are reported elsewhere
\cite{g-2,forth}. Here we are interested in the $M_{\pi\pi}$ region
around the $\eta$ mass where the signal is expected to be.
The $\eta$ mass region (500 - 600 MeV) of the $M_{\pi\pi}$ spectrum is
dominated by ISR events with the radiated photon
mostly at small polar angle. To reduce the amount of such events while keeping
a reasonable acceptance, we  require the photon to be at large
angle (45$\deg< \theta_{\gamma}< 135\deg$).

From Monte Carlo simulation
we find the overall signal efficiency to be,\\ 
\eps\dn s=(16.6\plm0.2\dn{\rm stat}\plm0.4\dn{syst})\%. The 2\% 
systematic uncertainty
is estimated by comparing the data and Monte Carlo distributions of the variables
$M_T$ and $\psi$. The overall rejection factors for the backgrounds range between
order 10$^4$ for $\mu^+\mu^-\gamma$ and 10$^6$ for $\pi^+\pi^-\pi^0$ and $e^+e^-\gamma$.

The expected $M_{\pi\pi}$ distribution for a possible signal is a Gaussian with a resolution of 1.33 MeV. 
Analysis of the similar and abundant decay
$K_s\rightarrow\pi^+\pi^-$ shows that the Monte Carlo  correctly reproduces the
observed mass distribution (see Ref. \cite{NIMoffline}).
\begin{figure}[ht]
\centering
\epsfig{file=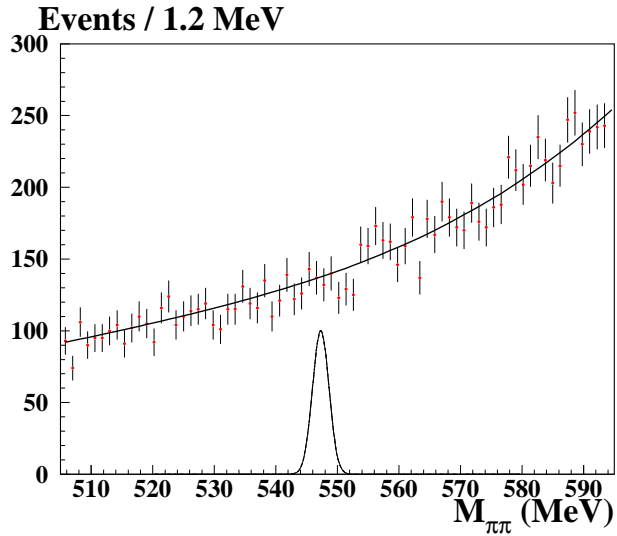,width=9cm}
\caption{$\pi^+\pi^-$ mass spectrum between 500 and 600 MeV 
superimposed to the fitted background. The Gaussian is the expected signal 
shape in arbitrary units.}
\label{unica}
\end{figure}
Figure \ref{unica} shows the measured $M_{\pi\pi}$ spectrum 
in the region around the $\eta$ mass, in 1.2 MeV 
bins, together with the expected form for the 
signal. No evidence for
the signal is
observed. The curve superimposed to the data is the result of a fit to the 
$M_{\pi\pi}$ spectrum over a much wider interval, from 410 MeV up to 1010 MeV
with a function
that describes all the appropriate physical processes:
\begin{equation}
{\dif N\over\dif M_{\pi\pi}}=\left({\dif\sigma(ISR)\over\dif M_{\pi\pi}} 
+{\dif\sigma(FSR+f_0)\over\dif M_{\pi\pi}}
+{\dif\sigma(\rho\pi)\over\dif M_{\pi\pi}} 
\right)\times {\mathcal L} \times \epsilon(M_{\pi\pi}).
\label{bckgnd}
\end{equation}
In Eq. \ref{bckgnd} above, ${\mathcal L}$ is the integrated luminosity of the 
sample analysed, $\epsilon(M_{\pi\pi})$ is 
the selection efficiency as a function
of $M_{\pi\pi}$ for the full range of the fit, and the 
$\dif\sigma/\dif M_{\pi\pi}$ terms
are the differential cross sections for the various processes which 
contribute to the background.
The fit has 7 free paramaters and gives a $\chi^2$ value of 75 for 
84 data points in the 500 - 600 MeV region and a value of 539 for 488 data points in the full mass
range.
4 out of the 7 parameters describe the pion form factor according
to Ref.\cite{Santamaria}: $M_{\rho}$, $\Gamma_{\rho}$, $\alpha$ and
$\beta$; the other 3 describe the scalar contribution according to
Ref.\cite{Achasov}: $M_{f_0}$, $g_{f_0KK}$ and $g_{f_0\pi\pi}$.
The result of this fit does not change if we remove the data points
in the signal region. We use this result as the estimate of the 
background magnitude in the following.

In order to determine an upper limit for the branching ratio, we have 
repeated the fit by adding to the previously estimated background a signal
component represented by a Gaussian with fixed mean and width of 547.3 and 1.33 MeV respectively, and free
absolute normalisation. The fit returns a number of signal events $N_s=-8\pm 24$, 
compatible with zero. 
The result does not depend on the choice of the fit interval and 
bin size.
The probability distribution of $N_s$ has been checked by generating
a large number of histograms according to the background
distribution and fitting each of them to get $N_s$. 
The results of this simple simulation show that $N_s$, in the 
case of no signal, 
is indeed Gaussian distributed with a mean compatible with zero and a width of
24.
The 90\% confidence-level upper limit on the number of events
is obtained using the tables in Ref. \cite{FeldCou}.
We find $N_s < 33$.

Alternatively, we have used a polynomial parametrisation of the
background, obtained by fitting the sideband regions (500 - 540 MeV and
555 - 600 MeV) only. Applying the same procedure to get $N_s$, we obtain  
$N_s=-10\pm 24$ and consequently $N_s < 31$. The systematic uncertainty due to
the parametrisation of the background is small and we use the largest value for
the limit.

The total number of $\eta$'s in the sample, $N_{\eta}$, is evaluated 
counting the number of $\phi\rightarrow\eta\gamma$ events
with $\eta\rightarrow 3\pi^0$.
The efficiency for this
channel is $\epsilon(\phi\rightarrow\eta\gamma,\eta\rightarrow
3\pi^0)=0.378\pm0.001_{stat}\pm0.008_{syst}$, where the 2\% systematic error
is dominated by the uncertainty on the detection efficiency of 
low energy photons.
Using the known branching fraction
BR($\eta\rightarrow 3\pi^0$)=0.3251$\pm$0.0029 \cite{PDG} we obtain:
\begin{equation}
N_{\eta}={{N(\eta\rightarrow 3\pi^0)}\over
{\epsilon(\phi\rightarrow\eta\gamma,\eta\rightarrow 3\pi^0) \times 
BR(\eta\rightarrow 3\pi^0)}}=1.55\times 10^7
\end{equation}
with a systematic uncertainty due to the knowledge of the 
efficiencies and of the
intermediate branching ratio of 2\%.

Taking the result $N_s < 33$, the 90\% confidence-level 
upper limit is
\begin{equation}
 BR(\eta\rightarrow\pi^+\pi^-) =
 {{N_s}\over{N_{\eta}\epsilon(\eta\rightarrow\pi^+\pi^-)}} < 1.3\times 10^{-5}.
\end{equation} 

This result is the best obtained to date and is 25 times more 
stringent than the previous best limit. 

\ack

We thank the DAFNE team for their efforts in maintaining low background running 
conditions and their collaboration during all data-taking. 
We want to thank our technical staff: 
G.F.Fortugno for his dedicated work to ensure an efficient operation of 
the KLOE Computing Center; 
M.Anelli for his continuos support to the gas system and the safety of the
detector; 
A.Balla, M.Gatta, G.Corradi and G.Papalino for the maintenance of the
electronics;
M.Santoni, G.Paoluzzi and R.Rosellini for the general support to the
detector; 
C.Piscitelli for
his help during major maintenance periods.
This work was supported in part by DOE grant DE-FG-02-97ER41027; 
by EURODAPHNE, contract FMRX-CT98-0169; 
by the German Federal Ministry of Education and Research (BMBF) contract 06-KA-957; 
by Graduiertenkolleg `H.E. Phys. and Part. Astrophys.' of Deutsche Forschungsgemeinschaft,
Contract No. GK 742; 
by INTAS, contracts 96-624, 99-37; 
and by TARI, contract HPRI-CT-1999-00088.

\end{document}